\documentclass[onecolumn,journal,12pt]{IEEEtran}
\usepackage[T1]{fontenc}
\usepackage{amsmath}
\usepackage{amsthm}
\usepackage{amssymb}
\usepackage{graphicx}
\usepackage{graphics}
\usepackage[unicode=true,
 bookmarks=true,bookmarksnumbered=true,bookmarksopen=true,bookmarksopenlevel=1,
 breaklinks=false,pdfborder={0 0 0},backref=false,colorlinks=false]
 {hyperref}
\hypersetup{pdftitle={Your Title},
 pdfauthor={Your Name},
 pdfpagelayout=OneColumn, pdfnewwindow=true, pdfstartview=XYZ, plainpages=false}
\usepackage{breakurl}

\makeatletter

\providecommand{\tabularnewline}{\\}

\theoremstyle{plain}
\newtheorem{thm}{\protect\theoremname}
\theoremstyle{remark}
\newtheorem{rem}[thm]{\protect\remarkname}
\theoremstyle{definition}
\newtheorem{defn}[thm]{\protect\definitionname}
\theoremstyle{definition}
\newtheorem{problem}[thm]{\protect\problemname}
\theoremstyle{plain}
\newtheorem{lem}[thm]{\protect\lemmaname}

\usepackage[caption=false,font=footnotesize]{subfig}

\makeatother

\providecommand{\definitionname}{Definition}
\providecommand{\lemmaname}{Lemma}
\providecommand{\problemname}{Problem}
\providecommand{\remarkname}{Remark}
\providecommand{\theoremname}{Theorem}

\begin{document}

\title{Robust Monotonic Convergent Iterative Learning Control Design: an LMI-based Method}

\author{Lanlan Su \thanks{
L. Su is with School of Engineering, University of Leicester, Leicester,
LE1 7RH, UK (email: ls499@leicester.ac.uk)}}

\maketitle
\begin{abstract}
This work investigates robust monotonic convergent iterative learning control (ILC) for uncertain
linear systems in both time and frequency domains, and the ILC algorithm
optimizing the convergence speed in terms of $l_{2}$ norm of error
signals is derived. Firstly, it is shown that the robust monotonic convergence of the ILC system can be established equivalently by  the positive definiteness of a matrix polynomial over some set. Then, a necessary and sufficient condition in the form of sum of squares (SOS) for the positive definiteness is proposed,  which is amendable to the feasibility of linear matrix inequalities (LMIs). Based on such a condition, the optimal ILC algorithm that maximizes the convergence speed is obtained by solving a set of convex optimization problems. Moreover, the order of the learning function can be chosen arbitrarily so that the designers have the flexibility to decide the complexity of the learning algorithm. \end{abstract}


\section{Introduction}

Iterative leaning control (ILC) is a useful control strategy to improve
tracking performance over repetitive trials (\cite{arimoto1984bettering,hara1988repetitive,moore2000special}). The basic idea
is to incorporate the error signals from previous iterations into
updating the control signal for subsequent iterations.
Stability is naturally the most fundamental topic in ILC which is widely
studied in both the time and frequency domain (see \cite{norrlof2002time} and the references therein for more details). However, in
some stable ILC systems the error can grow very large before converging
to the desired output trajectory, which is undesirable in most practical
applications \cite{ bristow2006survey,abidi2010iterative}. Hence, the objective
of monotone convergence is of crucial importance to improve tracking
performance from trial to trial (see, for instance \cite{moore2005monotonically,abidi2010iterative,seel2017monotonic}). 

A relatively new research line in the existing literature regarding ILC is robust monotone convergence in the presence of uncertain models. Several approaches have been proposed by researches for robust monotone convergence of  ILC. See, e.g., \cite{harte2005discrete} which proposes the inverse model-based ILC, \cite{ahn2006monotonic} which designs robust monotonic convergent ILC controller based on interval model conversion methods,  and the norm-optimal design with a quadratic cost function is considered in  \cite{son2015robust,ge2017frequency}. Since the $H_{\infty}$ method
offers a common approach to robust feedback controller design, there
exists a trend of applying it to robust ILC control. For instance, \cite{de1996synthesis}
studies the problem of robust convergent ILC by solving $\mu$-synthesis
problem;  \cite{van2009iterative} presents sufficient conditions for robust monotonic
convergence analysis based on $\mu$ analysis; \cite{moore2008iteration, zheng2017design} provide $H_{\infty}$-based
design method to synthesize high-order ILCs. It is generally known
that $H_{\infty}$-based robust controller computation  will
inevitably encounter the notoriously  difficulty caused by $\mu$-synthesis,
and consequently only sub-optimal controllers can be derived.

The abovementioned works provide  sufficient conditions for robust monotone convergence of ILC algorithms, and the ILC controller  that optimizes the robust monotone convergent rate can not be explicitly derived.  The optimal ILC algorithms  are proposed in  \cite{owens2009robust,de2020multivariable}. In \cite{owens2009robust}, a necessary and sufficient condition is proposed for the gradient-based ILC algorithm to be robustly monotonically convergent in time domain. \cite{de2020multivariable} studies the problem in frequency domain, and it shows that the optimal convergence rate can be achieved if the learning function is chosen to be the inverse nominal system. Compared to \cite{owens2009robust,de2020multivariable}, the form of the learning function of this work is more general in the sense that  the order of the learning function can be chosen arbitrarily so that the designers have the flexibility to decide the complexity of the learning algorithm.

This work investigates robust monotonic convergent ILC with learning function of flexible order for uncertain
linear systems in both time and frequency domains,  and the optimal
ILC algorithm which optimizes the convergence speed in terms of $l_{2}$
norm of error signals is derived. Specifically, we first show  that the robust monotonic convergence of the ILC system can be reformulated as  positive definiteness of a matrix polynomial over some set. Then we provide a necessary and sufficient condition for the  positive definiteness of the  matrix polynomial over the constrained set, which is equivalent to the feasibility of linear matrix inequalities (LMIs). Based on such a condition, the optimal ILC algorithm that maximizes the robust monotonic convergence speed can be obtained by solving a set of convex optimization problems. Moreover, the order of the learning function can be chosen arbitrarily in our results.  This enables the designers to decide the complexity of the learning algorithm based on the memory and computational
capability allowed by the computing platform.

\vspace{2mm}
The  paper is organized as follows.  In the rest of this section, the notation and the SOS techniques are introduced. In Section
2, the time domain results are presented while  Section 3 proposes the frequency domain results starting with the nominal case.  Section 4 provides an illustrative
example. Finally, Section 5 concludes the paper.

 \subsection*{Notation:} The sets of real numbers, complex numbers and nonnegative
integers are denoted by $\mathbb{R},\mathbb{C}$ and $\mathbb{N}$. The imaginary unit is denoted as $j$.
Let $\mathbb{R}^{n}$ denote the set of $n$-dimensional real vectors,
and the set $\mathbb{R}_{0}^{n}$ represents $\mathbb{R}^{n}\backslash\{0_{n}\}$.
Given a matrix $M$, its transpose and largest singular value are
denoted by $M^{T}$ and $\bar{\sigma}(M)$, and the notation $M>0$ (respectively, $M\ge 0$) denotes that $M$ is positive definite (respectively,  positive  semidefinite).  Denote by $I$ the identity
matrix with its dimension clear from the context.  For a vector $x=[x_{1},\ldots,x_{n}]^{T}$,
the notation $x^{2}$ denotes the vector of squares $[x_{1}^{2},\ldots,x_{n}^{2}]^{T}$,
and $x^{y}$ with $y\in\mathbb{N}^{n}$ denotes $\prod_{i=1}^{n}x_{i}^{y_{i}}$.
For a complex number $z\in\mathbb{C}$, its magnitude and conjugate
are denoted as $|z|$ and $\bar{z}$ respectively. The notation $\left\Vert \cdot\right\Vert _{2}$
denotes $l_{2}$ norm for signals, and we use it to also denote the
Euclidean norm for vectors when it is clear from the context. The
$H_{\infty}$ norm for systems is denoted by $\left\Vert \cdot\right\Vert _{\infty}$.
Given a matrix polynomial $P(x)$, the notation $\text{deg}(P)$ denotes
the maximum of the degrees of the entries of $P(x)$. For a matrix
polynomial in two sets of variables $P(x,y)$, we use $\text{deg}(P,x)$
(respectively, $\text{deg}(P,y)$) to denote the maximum degree of the entries of
$P$ in $x$ (respectively, $y$). The acronym SOS stands for sum of squares of
matrix polynomials.

\subsection*{Sum of square (SOS) matrix polynomial}\label{sos}

Next, we briefly introduce the class of SOS matrix polynomials, see
e.g., \cite{chesi2010lmi} for details.

A symmetric real matrix polynomial $F(s):\mathbb{R}^{r}\rightarrow\mathbb{R}^{n\times n}$
is said to be SOS if and only if there exist real matrix polynomials $F_{1}(s),\ldots,F_{k}(s):\mathbb{R}^{r}\rightarrow\mathbb{R}^{n\times n}$
such that $F(s)=\sum_{i=1}^{k}F_{i}(s)^{T}F_{i}(s).$ SOS matrix polynomials
are positive-semidefinite, i.e., $F(s)\ge 0$ for all $s\in\mathbb{R}^r$, and it turns out that one can establish
whether a symmetric matrix polynomial is SOS via an LMI feasibility
test. Indeed, let $d$ be a nonnegative integer such that $2d\ge\deg(F)$.
By extending the Gram matrix method, $F(s)$ can be written
as 
\begin{equation*}
F(s)=\left(b(s)\otimes I\right)^{T}\left(M+L(\alpha)\right)\left(b(s)\otimes I\right)\label{smrmat}
\end{equation*}
where $b(s):\mathbb{R}^{r}\rightarrow\mathbb{R}^{\sigma(r,d)}$ is
a vector containing all the monomials of degree less than or equal
to $d$ in $s$ with $\sigma(r,d)=\frac{(r+d)!}{r!d!}$ , and $M\in\mathbb{R}^{n\sigma(r,d)\times n\sigma(r,d)}$
is a symmetric matrix satisfying 
\[
F(s)=\left(b(s)\otimes I\right)^{T}M\left(b(s)\otimes I\right),
\]
 $L(\alpha):\mathbb{R}^{\omega(r,2d,n)}\rightarrow\mathbb{R}^{n\sigma(r,d)\times n\sigma(r,d)}$
is a linear parametrization of the linear set 
\begin{equation*}
\mathcal{L}=\{\tilde{L}=\tilde{L}^{T}:~\left(b(s)\otimes I\right)^{T}\tilde{L}\left(b(s)\otimes I\right)=0\},\label{L}
\end{equation*}
and $\alpha\in\mathbb{R}^{\omega(r,2d,n)}$ is a free vector with
$\omega(r,2d,n)=\frac{1}{2}n\left(\sigma(r,d)(n\sigma(r,d)+1)-(n+1)\sigma(r,2d)\right).$ 
It follows that $F(s)$ is a SOS matrix polynomial if and only if
there exists $\alpha$ satisfying the LMI 
\[
M+L(\alpha)\ge0.
\]

\section{Time-domain results with uncertain plant}

In this section, we focus on robust ILC design with finite-time interval
per trial denoted as $N$. 

Consider the uncertain plant 
\begin{equation}
P(q,\lambda)=p_{1}(\mathbf{\lambda})q^{-1}+p_{2}(\lambda)q^{-2}+\cdots p_{N}(\lambda)q^{-N},\forall\lambda\in\Lambda\label{eq: uncertain plant}
\end{equation}
where the Markov parameters  $p_{1},\ldots,p_{N}$ depend rationally on the
uncertainty vector $\lambda\in\Lambda$ with $p_{1}\neq0,\forall\lambda\in\Lambda$,
and $q$ is the forward time-shift operator. The Markov parameters can be obtained from  uncertain linear time invariant systems represented by state-space equations or transfer functions. 

We suppose that $\lambda\in\mathbb{R}^{n}$ is constrained into a
simplex $\Lambda$, which is defined as follows
\[
\Lambda=\left\{ \lambda\in\mathbb{R}^{n}:\ \sum_{i=1}^{n}\lambda_{i}=1,\lambda_{i}\ge0,\forall i=1,\ldots,n\right\} .
\]

\begin{rem}
\label{rem:1}We note that the formulation of the uncertain plant
\eqref{eq: uncertain plant} can be equivalently reformulated into
the case where the uncertainty vector is constrained into a generic
bounded convex polytope. In fact, consider the plant described as
\begin{equation}
P(q,\text{\ensuremath{\theta}})=p_{1}(\theta)q^{-1}+p_{2}(\theta)q^{-2}+\cdots p_{N}(\theta)q^{-N},\forall\theta\in\Theta\label{eq: uncertain plant 2}
\end{equation}
where $p_{1},\ldots,p_{N}$ are rational function in the uncertainty
vector $\theta$, and $\theta$ belongs to a bounded convex polytope
$\Theta$. Then the vector $\theta$ in $\Theta$ can be replaced
by a linear function $l(\lambda)$ over $\Lambda$ by the change of
variable $\theta=l(\lambda).$ Hence, by substituting $\theta$ with
$l(\lambda)$, \eqref{eq: uncertain plant 2} can be rewritten as in
\eqref{eq: uncertain plant}. It should be mentioned that the formulation
\eqref{eq: uncertain plant} has also included the uncertain interval
plant model represented by 
\[
\begin{array}{c}
P(q)=p_{1}q^{-1}+p_{2}q^{-2}+\ldots+p_{N}q^{-N}\\
p_{i}\in[\underline{p}_{i},\bar{p}_{i}],\ i=1,\ldots,N
\end{array}
\]
as a special case, as the uncertain coefficients $p_{1},p_{2},\ldots,p_{N}$
are constrained into a hyper-rectangle which is a bounded convex polytope. 
\end{rem}
The lifted system representation is given by
\begin{equation}
\underset{\mathbf{y_{j}}}{\underbrace{\begin{pmatrix}y_{j}(1)\\
y_{j}(2)\\
\vdots\\
y_{j}(N)
\end{pmatrix}}}=\underset{\mathbf{P(\lambda)}}{\underbrace{\begin{pmatrix}p_{1}(\lambda) & 0 & \cdots & 0\\
p_{2}(\lambda) & p_{1}(\lambda) & \cdots & 0\\
\vdots & \vdots & \ddots & \vdots\\
p_{N}(\lambda) & p_{N-1}(\lambda) & \cdots & p_{1}(\lambda)
\end{pmatrix}}}\underset{\mathbf{u_{j}}}{\underbrace{\begin{pmatrix}u_{j}(0)\\
u_{j}(1)\\
\vdots\\
u_{j}(N-1)
\end{pmatrix}}}+\underset{\mathbf{d}}{\underbrace{\begin{pmatrix}d(1)\\
d(2)\\
\vdots\\
d(N)
\end{pmatrix}}}\label{eq:uncertain plant 3}
\end{equation}
where $\mathbf{y_{j},u_{j}}$ denote the $N$-dimensional vectors
containing output and input signals during the $j$-th trial, respectively,
and $\mathbf{d}$ is an exogenous signal which repeats at each trial. 

We denote by $\mathbf{y_{d}}\in\mathbb{R}^{N}$ the desired output
trajectory, and the output error of the $j$-th trial, $\mathbf{e_{j}},$
is defined as
\begin{equation}
\underset{\mathbf{e_{j}}}{\underbrace{\begin{pmatrix}e_{j}(1)\\
e_{j}(2)\\
\vdots\\
e_{j}(N)
\end{pmatrix}}}=\underset{\mathbf{y_{d}}}{\underbrace{\begin{pmatrix}y_{d}(1)\\
y_{d}(2)\\
\vdots\\
y_{d}(N)
\end{pmatrix}}}-\underset{\mathbf{y_{j}}}{\underbrace{\begin{pmatrix}y_{j}(1)\\
y_{j}(2)\\
\vdots\\
y_{j}(N)
\end{pmatrix}}}.\label{eq:error}
\end{equation}
We adopt the widely used ILC learning algorithm  as follows \cite{bristow2006survey, bristow2008monotonic}:
\begin{equation}
u_{j+1}(k)=Q(q)\left[u_{j}(k)+L(q)e_{j}(k+1)\right]\label{eq: learning in time}
\end{equation}
wherein the Q-filter $Q(q)$ and learning function $L(q)$ are allowed
to be non-causal, described respectively by 
\begin{equation}
Q(q)=q_{-(N-1)}q^{N-1}+\cdots+q_{-1}q^{1}+q_{0}+q_{1}q^{-1}+\cdots+q_{N-1}q^{-(N-1)}\label{eq: Q(q)}
\end{equation}
and 
\begin{equation}
L(q)=l_{-(N-1)}q^{N-1}+\cdots+l_{-1}q^{1}+l_{0}+l_{1}q^{-1}+\cdots+l_{N-1}q^{-(N-1)}\label{eq: L(q)}
\end{equation}
For brevity, denote the vector $(l_{-(N-1)},\cdots,l_{-1},l_0,l_1,\cdots,l_{N-1})^{T}$ as $l$. In this work, the coefficients in \eqref{eq: Q(q)} and \eqref{eq: L(q)}  can be fixed as zeros or  other constants required by the designer, and the order of the ILC algorithm can be reduced. This enables the designers to decide the complexity of the learning algorithm based on the memory and computational capability allowed by the computing platform.

It is shown in \cite{phan2000unified} that the identity Q-filter $Q(q)=1$ is necessary for $\mathbf{e}_{\infty}=0$.  While most existing works do not include the Q-filter, i.e., $Q(q)=1$,  for ensuring perfect tracking, the incorporation of the Q-filter enables us to improve transient learning behavior and robustness (see Example 1 in \cite{bristow2006survey} for instance). More insights of Q-filter can be found in  \cite{de1996synthesis}. Without loss of generality, we include Q-filter in this work.

Equations \eqref{eq: learning in time}-\eqref{eq: L(q)} can be rewritten
into the lifted form: 
\begin{equation}
\begin{array}{rl}
\underset{\mathbf{u_{j+1}}}{\underbrace{\begin{pmatrix}u_{j+1}(0)\\
u_{j+2}(1)\\
\vdots\\
u_{j+1}(N-1)
\end{pmatrix}}}= & \underset{\mathbf{Q}}{\underbrace{\begin{pmatrix}q_{0} & q_{-1} & \cdots & q_{-(N-1)}\\
q_{1} & q_{0} & \cdots & q_{-(N-2)}\\
\vdots & \vdots & \ddots & \vdots\\
q_{N-1} & q_{N-2} & \cdots & q_{0}
\end{pmatrix}}}\underset{\mathbf{u_{j}}}{\left\{ \underbrace{\begin{pmatrix}u_{j}(0)\\
u_{j}(1)\\
\vdots\\
u_{j}(N-1)
\end{pmatrix}}\right.}\\
 & +\left.\underset{\mathbf{L}}{\underbrace{\begin{pmatrix}l_{0} & l_{-1} & \cdots & l_{-(N-1)}\\
l_{1} & l_{0} & \cdots & l_{-(N-2)}\\
\vdots & \vdots & \ddots & \vdots\\
l_{N-1} & l_{N-2} & \cdots & l_{0}
\end{pmatrix}}}\underset{\mathbf{e_{j}}}{\underbrace{\begin{pmatrix}e_{j}(1)\\
e_{j}(2)\\
\vdots\\
e_{j}(N)
\end{pmatrix}}}\right\} 
\end{array}\label{eq:ILC controller}
\end{equation}
 In this section we aim at designing the learning algorithm \eqref{eq:ILC controller}
to ensure the robust monotonic convergence defined as follows.
\begin{defn}
\textbf{\label{def:(Robust-Monotonic-Convergence)}(Robust Monotonic
Convergence)} The uncertain system \eqref{eq:uncertain plant 3} with
the ILC algorithm \eqref{eq:ILC controller} is robustly monotonically
convergent if there exists $0\le\gamma<1$ such that 
\[
\left\Vert \mathbf{e}_{\infty}-\mathbf{\mathbf{e}}_{j+1}\right\Vert _{2}<\gamma\left\Vert \mathbf{e}_{\infty}-\mathbf{\mathbf{e}}_{j}\right\Vert _{2},\forall j\in\mathbb{N},\forall\lambda\in\Lambda
\]
where $\gamma$ is called the convergence rate.
\end{defn}
From the ILC system dynamics \eqref{eq:uncertain plant 3}, \eqref{eq:ILC controller}
and the error \eqref{eq:error}, it can be derived by some algebraic
manipulations that 
\[
\mathbf{e}_{\infty}-\mathbf{\mathbf{e}}_{j+1}=\mathbf{P}(\lambda)\mathbf{Q}(I-\mathbf{LP}(\lambda))\mathbf{P}^{-1}(\lambda)(\mathbf{e}_{\infty}-\mathbf{\mathbf{e}}_{j}).
\]
Hence, the uncertain system \eqref{eq:uncertain plant 3} with the
ILC algorithm \eqref{eq:ILC controller} is robustly monotonically
convergent if and only if 
\begin{equation}
\gamma\triangleq\underset{\lambda\in\Lambda}{\text{max}}\;\bar{\sigma}\left(\mathbf{P}(\lambda)\mathbf{Q}(I-\mathbf{LP}(\lambda))\mathbf{P}^{-1}(\lambda)\right)<1.\label{eq:def gamma}
\end{equation}

\begin{problem}
\label{prob:1}Given a fixed $\mathbf{Q}$, find the learning matrix
$\mathbf{L}$ such that the convergent rate $\gamma$ defined in \eqref{eq:def gamma}
is minimized, i.e., 
\begin{equation}
\begin{array}{c}
\underset{\mathbf{L}}{\text{min}} \; \underset{\lambda\in\Lambda}{\text{sup}} \;\bar{\sigma}\left(\mathbf{P}(\lambda)\mathbf{Q}(I-\mathbf{LP}(\lambda))\mathbf{P}^{-1}(\lambda)\right).\end{array}\label{eq:prob1}
\end{equation}

Let us observe that the optimization problem in \eqref{eq:prob1}
can be transformed into minimizing $\gamma$ over $\mathbf{L}$ subject to
\begin{equation}
\begin{array}{r}
\gamma^{2}I-\left(\mathbf{P}(\lambda)\mathbf{Q}(I-\mathbf{LP}(\lambda))\mathbf{P}^{-1}(\lambda)\right)^{T}\cdot\\
\left(\mathbf{P}(\lambda)\mathbf{Q}(I-\mathbf{LP}(\lambda))\mathbf{P}^{-1}(\lambda)\right)>0,\forall\lambda\in\Lambda.
\end{array}\label{eq:var 1}
\end{equation}
Now let us express 
\[
\mathbf{P}(\lambda)\mathbf{Q}(I-\mathbf{LP}(\lambda))\mathbf{P}^{-1}(\lambda)=\frac{W(L,\lambda)}{a(\lambda)}
\]
where $a(\lambda)=\det(\mathbf{P}(\lambda))$ and $W(L,\lambda)=\mathbf{P}(\lambda)\mathbf{Q}(I-\mathbf{LP}(\lambda))\text{adj}(\mathbf{P}(\lambda))$.
Then, the constraint \eqref{eq:var 1} transforms to 
\[
\gamma^{2}a^{2}(\lambda)I-W^{T}(L,\lambda)W(L,\lambda)>0,\forall\lambda\in\Lambda
\]
which can be rewritten into 
\begin{equation}
\begin{pmatrix}\gamma^{2}a^{2}(\lambda)I & W^{T}(L,\lambda)\\
W(L,\lambda) & I
\end{pmatrix}>0,\forall\lambda\in\Lambda.\label{eq:var 2}
\end{equation}

Before continuing, we denote the operator that returns a matrix homogeneous polynomial $\bar{G}(\lambda)$
  from a matrix polynomial $G(\lambda)$
  in the variable $\lambda\in\Lambda$
  while satisfying $\bar{G}(\lambda)=G(\lambda),\forall\lambda\in\Lambda$
  as
\begin{equation}
\bar{G}(\lambda)=\text{hom}(G(\lambda),\lambda).\label{eq:hom}
\end{equation}
 Since $\sum_{i=1}^{n}\lambda_{i}=1$, such an operation can be done by simply multiplying each monomial of $G(\lambda)$ by a suitable power of $\sum_{i=1}^{n}\lambda_{i}$.

Now we are ready to define
\begin{equation}
M(\gamma^{2},L,\lambda)=\text{hom}\left(\begin{pmatrix}\gamma^{2}a^{2}(\lambda)I & W^{T}(L,\lambda)\\
W(L,\lambda) & I
\end{pmatrix},\lambda\right)\label{eq:hom 1}
\end{equation}
and denote by $\text{deg}(M)$ the degree of matrix polynomial $M$
with respect to $\lambda$. \end{problem}
\begin{lem}
\label{lem:1}The condition \eqref{eq:var 2} holds if and only if
there exists a scalar $\epsilon>0$ and an integer $k\in\mathbb{N}$
such that 
\begin{equation}
\left(M(\gamma^{2},L,\lambda^{2})-\epsilon\left\Vert \lambda\right\Vert _{2}^{2\text{deg}(M)}\right)\left\Vert \lambda\right\Vert _{2}^{2k}\ \ \text{is SOS. }\label{eq: sos condition 1}
\end{equation}
\end{lem}
\begin{IEEEproof}
``$\Leftarrow$'' Suppose there exists a scalar $\epsilon>0$ and
an integer $k\in\mathbb{N}$ satisfying \eqref{eq: sos condition 1}.
It follows that 
\[
M(\gamma^{2},L,\lambda^{2})>0,\forall\lambda\in\mathbb{R}_{0}^{n}.
\]
By exploiting Theorem 11 in \cite{chesi2010lmi}, we have that $M(\gamma^{2},L,\lambda)>0,\forall\lambda\in\Lambda$.
This is the same with \eqref{eq:var 2} due to 
\[
M(\gamma^{2},L,\lambda)=\begin{pmatrix}\gamma^{2}a^{2}(\lambda)I & W^{T}(L,\lambda)\\
W(L,\lambda) & I
\end{pmatrix},\forall\lambda\in\Lambda.
\]

``$\Rightarrow$'' Now assume the condition \eqref{eq:var 2} holds.
Since $\Lambda$ is a compact, there exists a small enough scalar
$\epsilon>0$ such that 
\[
\begin{pmatrix}\gamma^{2}a^{2}(\lambda)I & W^{T}(L,\lambda)\\
W(L,\lambda) & I
\end{pmatrix}-\epsilon I>0,\forall\lambda\in\Lambda,
\]
which is equivalent to $M(\gamma^{2},L,\lambda)-\epsilon\left(\sum_{i=1}^{n}\lambda_{i}\right)^{\text{deg}(M)}I>0,\forall\lambda\in\Lambda.$
Next, let us apply the extended Polya's Theorem 3 in \cite{scherer2006matrix},
and obtain that there exists $k\in\mathbb{N}$ such that all coefficients
of $\left[M(\gamma^{2},L,\lambda)-\epsilon\left(\sum_{i=1}^{n}\lambda_{i}\right)^{\text{deg}(M)}I\right]\left(\sum_{i=1}^{n}\lambda_{i}\right)^{k}$
with respect to $\lambda$ are positive definite. Rewrite 
\[
\left[M(\gamma^{2},L,\lambda)-\epsilon\left(\sum_{i=1}^{n}\lambda_{i}\right)^{\text{deg}(M)}I\right]\left(\sum_{i=1}^{n}\lambda_{i}\right)^{k}=\sum_{\xi\in\Xi}C_{\xi}\lambda^{\xi}
\]
where $\Xi=\left\{ \xi\in\mathbb{N}^{n}:\sum_{i=1}^{n}\xi_{i}=\text{deg}(M)+k\right\} $,
and it follows that all $C_{\xi}>0.$ Hence, we can express $C_{\xi}$
as $C_{\xi}=D_{\xi}^{T}D_{\xi}$ by Cholesky decomposition. Now let
us replace $\lambda$ with $\lambda^{2},$ and it can be obtained
that 
\[
\begin{array}{rl}
 & \left(M(\gamma^{2},L,\lambda^{2})-\epsilon\left\Vert \lambda\right\Vert _{2}^{2\text{deg}(M)}\right)\left\Vert \lambda\right\Vert _{2}^{2k}\\
= & \sum_{\xi\in\Xi}\left(D_{\xi}\lambda^{\xi}\right)^{T}\left(D_{\xi}\lambda^{\xi}\right),
\end{array}
\]
which shows that the condition \eqref{eq: sos condition 1} holds.\end{IEEEproof}
\begin{rem}
\label{rem:2}It is worth mentioning that the condition \eqref{eq: sos condition 1}
is necessary when $k$ is sufficiently large. In fact, the upper bounds
of $k$ required for achieving necessity in Lemma \ref{lem:1} has
been investigated in \cite{scherer2006matrix}. \end{rem}
\begin{thm}
\label{thm:1}The optimization problem \eqref{eq:prob1} can be solved
by the following SOS program:

\begin{equation}
\begin{array}{rl}
\underset{{\scriptstyle \begin{array}{c}
\eta>0,\epsilon>0\\
k\in\mathbb{N},l
\end{array}}}{\text{min}} & \eta\\
\text{s.t.} & \left(M(\eta,L,\lambda^{2})-\epsilon\left\Vert \lambda\right\Vert _{2}^{2\text{deg}(M)}\right)\left\Vert \lambda\right\Vert_{2}^{2k}\ \ \text{is SOS }
\end{array}\label{eq:thm 1}
\end{equation}
where $\gamma^{*}=\sqrt{\eta^{*}}$ and $l^{*}$ is given by the optimal
solution of \eqref{eq:thm 1}.\end{thm}
\begin{rem}
\label{rem:3}Theorem \ref{thm:1} provides an approach to address
Problem \ref{prob:1} via solving convex optimization problems.
Specifically, for fixed $k\in\mathbb{N},$ the condition in \eqref{eq:thm 1}
can be verified through an SDP for the reason that the condition for
a matrix polynomial depending linearly on some decision variables
to be SOS is equivalent to an LMI feasibility test \cite{chesi2010lmi}. For each fixed $k$, the SOS program in Theorem \ref{thm:1} provides an upper bound for $\gamma ^*$, and this upper bound becomes strict when $k$ is sufficiently large. To sum up,
to solve \eqref{eq:thm 1}, one can start solving the SDP with small
fixed degree $k$, and then repeat for increased $k$ until the optimal
solution $\eta^{*}$ converges. Observe that $\eta^{*}$
will decrease as $k$ increases, and Theorem \ref{thm:1} is guaranteed
to be nonconservative for sufficiently large $k$ based on Remark
\ref{rem:2}. 
\end{rem}

\section{z-domain results }

When the interval per trial $N$ is large, it is obvious that the
matrix $\mathbf{P}(\lambda)$ in \eqref{eq:uncertain plant 3} as
well as the associating matrix $M$ in \eqref{eq:hom 1} will have
large size. This will inevitably lead to high computation complexity
making the optimization in Theorem \ref{thm:1} much less tractable.
To circumvent that, we resort to the frequency-domain approach by
assuming infinite trial length. \footnote{This is a standard assumption in ILC when frequency domain analysis
methods are exploited. See \cite{amann1996h} for more details. }

In this section, we consider the uncertain plant represented in $z$-domain
as 
\begin{equation}
P(z,\lambda)=\frac{b_{m}(\lambda)z^{m}+b_{m-1}(\lambda)z^{m-1}+\cdots+b_{0}(\lambda)}{z^{n}+a_{n-1}(\lambda)z^{n-1}+\cdots+a_{0}(\lambda)}\label{eq:plant_tf_uncertain}
\end{equation}
where $\lambda$ is the uncertainty vector. To make the frequency
domain representation well defined, we assume that $P(z,\lambda)$
are stable for all $\lambda$ constrained in some set. This can be
verified by various methods, see \cite{su2017design,su2016robust} for instance.
When the trial length $N$ is large, the system \eqref{eq: uncertain plant}  can be approximately described  in z domain as 
\begin{equation}
Y_{j}(z)=P(z,\lambda)U_{j}(z)+D(z)\label{eq:system}
\end{equation}
where $P(z,\lambda)$ is simply  $P(q,\lambda)$ with $q$ replaced by $z$. 
The learning algorithm \eqref{eq: learning in time} is transformed
to 
\begin{equation}
U_{j+1}(z)=Q(z)\left[U_{j}(z)+zL(z)E_{j}(z)\right]\label{eq: learning algorithm}
\end{equation}
where $E_{j}(z)=Y_{d}(z)-Y_{j}(z)$. Note that the combined effects of the initial  
condition  of the plant and the disturbance  can be reflected in $D(z)$. 

Without loss of generality, we consider non-causal Q-filter $Q(z)$
and learning function $L(z)$ with finite impulse responses
\begin{equation}
L(z)=l_{-k_{1}}z^{k_{1}}+\cdots+l_{-1}z+l_{0}+l_{1}z^{-1}+\cdots l_{-k_{2}}z^{-k_{2}}\label{eq:L(z)}
\end{equation}
\begin{equation}
Q(z)=q_{-k_{3}}z^{k_{3}}+\cdots+q_{-1}z+q_{0}+q_{1}z^{-1}+\cdots q_{-k_{4}}z^{-k_{4}}.
\end{equation}
 As it will be shown later,  the subsequent results are  applicable to arbitrarily chosen  $k_{1},k_{2},k_{3},k_{4}\in\mathbb{N}$. This enables the designers to decide the complexity of the learning algorithm based on the memory and computational capability allowed by the computing platform. 

As in the previous section, we aim at achieving the robust monotonic convergence. To this end, observe that
the error dynamics can be obtained as
\[
E_{\infty}(z)-E_{j+1}(z)=Q(z)\left[1-zL(z)P(z,\lambda)\right]\left(E_{\infty}(z)-E_{j}(z)\right).
\]

Analogous to Definition \ref{def:(Robust-Monotonic-Convergence)},
the uncertain system \eqref{eq:system} with the ILC algorithm \eqref{eq: learning algorithm}
is said to be robustly monotonically convergent if there exists $0\le\gamma<1$
such that 
\[
\left\Vert E_{\infty}(z)-E_{j+1}(z)\right\Vert _{2}<\gamma\left\Vert E_{\infty}(z)-E_{j}(z)\right\Vert _{2},\ \forall j\in\mathbb{N},\forall\lambda\in\Lambda.
\]
This is equivalent to 
\begin{equation}
\gamma\triangleq\left\Vert Q(z)\left[1-zL(z)P(z,\lambda)\right]\right\Vert _{\infty}<1,\forall\lambda\in\Lambda.\label{eq:gamma_z-domain}
\end{equation}

 The problem addressed in this section is as follows. 
\begin{problem}
\label{prob:2}Given a fixed $Q(z)$, find the learning function $L(z)$
such that the convergent rate $\gamma$ defined in \eqref{eq:gamma_z-domain}
is minimized, i.e., 
\begin{equation}
\begin{array}{c}
\underset{L}{\text{min}} \; \underset{\lambda\in\Lambda}{\text{sup}}\left\Vert Q(z)\left[1-zL(z)P(z,\lambda)\right]\right\Vert _{\infty}.\end{array}\label{eq: prob 2}
\end{equation}

\end{problem}

\subsection{Nominal case}

Let us start by considering the nominal case with the plant given
by a rational transfer function 
\begin{equation}
P(z)=\frac{b_{m}z^{m}+b_{m-1}z^{m-1}+\cdots+b_{0}}{z^{n}+a_{n-1}z^{n-1}+\cdots+a_{0}}.\label{eq:plant_tf}
\end{equation}

The optimization problem in \eqref{eq: prob 2} is now simplified as
\begin{equation}
\begin{array}{rl}
\underset{\gamma,l}{\text{min}} & \gamma\\
 & \gamma^{2}-\overline{Q(z)\left[1-zL(z)P(z)\right]}Q(z)\left[1-zL(z)P(z)\right]>0\\
 & \forall z\in\mathbb{C},\left|z\right|=1
\end{array}\label{eq:optimization 1_z domain}
\end{equation}
where $l$ is the vector containing $l_{-k_{1}},\ldots,l_{0},\ldots,l_{k_{2}}$
to be optimized. Since $z$ is constrained into a compact set, one
has that the constraint in \eqref{eq:optimization 1_z domain} holds
if and only if there exists a scalar $\epsilon>0$ satisfying 
\begin{equation}
\gamma^{2}-\overline{Q(z)\left[1-zL(z)P(z)\right]}Q(z)\left[1-zL(z)P(z)\right]-\epsilon\ge0,\forall z\in\mathbb{C},\left|z\right|=1.\label{eq:optimization 1_z domain_1}
\end{equation}

Let $x\in\mathbb{R}$ be a new variable and define the function $\phi:\mathbb{R}\rightarrow\mathbb{C}$
as $\phi(x)=\frac{1-x^{2}+j2x}{1+x^{2}}$. It can be observed that the
complex unit circle $|z|=1$ can be parameterized by the real variable
$x\in\mathbb{R}.$ As a result, the optimization problem \eqref{eq:optimization 1_z domain}
can be equivalently reformulated as 

\begin{equation}
\begin{array}{rl}
\underset{\epsilon>0,\gamma,l}{\text{min}} & \gamma\\
 & \left(\gamma^{2}-\epsilon\right)-\overline{Q(\phi(x))\left[1-\phi(x)L(\phi(x))P(\phi(x))\right]}\cdot\\
 & \ \ \ \ \ \ \ \ Q(\phi(x))\left[1-\phi(x)L(\phi(x))P(\phi(x))\right]\ge0,\forall x\in\mathbb{R}.
\end{array}\label{eq:optimization_2 z-domain}
\end{equation}
Next, let us express 
\begin{equation}
\begin{array}{ll}
 & Q(\phi(x))\left[1-\phi(x)L(\phi(x))P(\phi(x))\right]\\
= & \left({\displaystyle \sum_{i=-k_{3}}^{k_{4}}}q_{i}\left(\frac{1-x^{2}+j2x}{1+x^{2}}\right)^{-i}\right)\left[1-\frac{1-x^{2}+j2x}{1+x^{2}}\left({\displaystyle \sum_{i=-k_{1}}^{k_{2}}}l_{i}\right.\right.\\
 & \left.\left.\left(\frac{1-x^{2}+j2x}{1+x^{2}}\right)^{-i}\right)\left(\frac{\sum_{i=0}^{m}b_{i}\left(\frac{1-x^{2}+j2x}{1+x^{2}}\right)^{i}}{\left(\frac{1-x^{2}+j2x}{1+x^{2}}\right)^{n}+\sum_{i=0}^{n-1}a_{i}\left(\frac{1-x^{2}+j2x}{1+x^{2}}\right)^{i}}\right)\right]\\
= & \frac{\tau_{1}(l,x)+j\tau_{2}(l,x)}{\tau_{3}(x)}
\end{array}\label{eq:transform}
\end{equation}
where $\tau_{1}$ and $\tau_{2}$ are polynomial in $x$ with their
coefficients depending linearly on $l$. 
\begin{lem}
The optimal solution $(\gamma^{*},l^{*})$ to \eqref{eq:optimization 1_z domain}
can be obtained by solving the following SOS program

\begin{equation}
\begin{array}{rl}
\underset{\eta>0,\epsilon>0,l}{\text{min}} & \eta\\
\text{s.t. } & \begin{pmatrix}\begin{array}{llc}
\left(\eta-\epsilon\right)\tau_{3}(x)^{2} & \tau_{1}(l,x) & \tau_{2}(l,x)\\
\tau_{1}(l,x) & 1 & 0\\
\tau_{2}(l,x) & 0 & 1
\end{array}\end{pmatrix}\text{is SOS}
\end{array}\label{eq: optimization z-domain SOS}
\end{equation}
where $\gamma^{*}=\sqrt{\eta^{*}}$ and $l^{*}$ is given by the optimal
solution. \end{lem}
\begin{IEEEproof}
By \eqref{eq:transform}, the constraint in \eqref{eq:optimization_2 z-domain}
can be rewritten into 
\[
\left(\gamma^{2}-\epsilon\right)\tau_{3}(x)^{2}-\tau_{1}(l,x)^{2}-\tau_{2}(l,x)^{2}\ge0,
\]
which is equivalent to 
\[
\begin{pmatrix}\begin{array}{llc}
\left(\gamma^{2}-\epsilon\right)\tau_{3}(x)^{2} & \tau_{1}(l,x) & \tau_{2}(l,x)\\
\tau_{1}(l,x) & 1 & 0\\
\tau_{2}(l,x) & 0 & 1
\end{array}\end{pmatrix} \ge 0
\]
according to the Schur complement lemma. 

From Theorem 4 in \cite{chesi2010lmi}, one has that a necessary and sufficient
condition for a univariate matrix polynomial being positive semidefinite
is that it is an SOS matrix. Consequently, the above constraint can
be equivalently expressed as 
\[
\begin{pmatrix}\begin{array}{llc}
\left(\eta-\epsilon\right)\tau_{3}(x)^{2} & \tau_{1}(l,x) & \tau_{2}(l,x)\\
\tau_{1}(l,x) & 1 & 0\\
\tau_{2}(l,x) & 0 & 1
\end{array}\end{pmatrix}\text{ is SOS}.
\]

Therefore, addressing \eqref{eq: optimization z-domain SOS} equates
to solving \eqref{eq:optimization 1_z domain} with $\gamma=\eta^{2}$. 
\end{IEEEproof}

\subsection{Uncertain case}

In this rest of this section, we proceed to the case whereas the plant
is affected by uncertainty. 

Consider the uncertain plant \eqref{eq: uncertain plant} with $a_{n},a_{n-1},\ldots,a_{0}$
and $b_{m},\ldots,b_{0}$ allowed to be any rational functions in
the uncertainty vector $\lambda.$ We assume that the uncertainty vector
$\lambda\in\mathbb{R}^{n}$ is constrained into a simplex $\Lambda$. 

The optimization problem \eqref{eq: prob 2} is now given as
\begin{equation}
\begin{array}{rl}
\underset{\gamma,l}{\text{min}} & \gamma\\
 & \gamma^{2}-\overline{Q(z)\left[1-zL(z)P(z,\lambda)\right]}\cdot\\
 & \ \ \ \ \ \ \ \ \ \ Q(z)\left[1-zL(z)P(z,\lambda)\right]>0\\
 & \forall z\in\mathbb{C},\left|z\right|=1,\forall\lambda\in\Lambda.
\end{array}\label{eq:optimzation_Z_uncertain_1}
\end{equation}

Let $x_{1},x_{2}\in\mathbb{R}$ be new variables. By substituting
$z$ with $\varphi(x_{1},x_{2})=x_{1}+jx_{2}$, one can observe that
$\left|z\right|=1$ can be parameterized by $x_{1},x_{2}\in\mathbb{R}$
subject to $x_{1}^{2}+x_{2}^{2}=1.$ Then \eqref{eq:optimzation_Z_uncertain_1}
can be transformed to 

\begin{equation}
\begin{array}{c}
\gamma^{2}-\overline{Q(\varphi(x_{1},x_{2}))\left[1-\varphi(x_{1},x_{2})L(\varphi(x_{1},x_{2}))P(\varphi(x_{1},x_{2}),\lambda)\right]}\\
\cdot Q(\varphi(x_{1},x_{2}))\left[1-\varphi(x_{1},x_{2})L(\varphi(x_{1},x_{2}))P(\varphi(x_{1},x_{2}),\lambda)\right]>0\\
\forall x_{1},x_{2}\in\mathbb{R},x_{1}^{2}+x_{2}^{2}=1,\forall\lambda\in\Lambda.
\end{array}\label{eq:transform-1}
\end{equation}

Similar to \eqref{eq:transform}, we can express 
\begin{equation}
\begin{array}{ll}
 & Q(\varphi(x_{1},x_{2}))\left[1-\varphi(x_{1},x_{2})L(\varphi(x_{1},x_{2}))P(\varphi(x_{1},x_{2}),\lambda)\right]\\
= & \frac{\nu_{1}(l,x_{1},x_{2},\lambda)+j\nu_{2}(l,x_{1},x_{2},\lambda)}{\nu_{3}(x_{1},x_{2},\lambda)}
\end{array}\label{eq:transform-2}
\end{equation}
where $\nu_{1}$ and $\nu_{2}$ are polynomials in $\left(x_{1},x_{2},\lambda\right)$
with their coefficients depending linearly on $l.$ 

Define 
\begin{equation}
\begin{array}{l}
\ \ \ \ \ T(l,\gamma^{2},x_{1},x_{2},\lambda)\\
=\begin{pmatrix}\begin{array}{ccc}
\gamma^{2}\nu_{3}(\lambda,x_{1},x_{2})^{2} & \nu_{1}(l,\lambda,x_{1},x_{2}) & \nu_{2}(l,\lambda,x_{1},x_{2})\\
\nu_{1}(l,\lambda,x_{1},x_{2}) & 1 & 0\\
\nu_{2}(l,\lambda,x_{1},x_{2}) & 0 & 1
\end{array}\end{pmatrix}
\end{array}\label{eq:T}
\end{equation}
and let 
\begin{equation}
\bar{T}(l,\gamma^{2},x_{1},x_{2},\lambda)=\text{hom}\left(T(l,\gamma^{2},x_{1},x_{2},\lambda),\lambda\right)\label{eq:hom2}
\end{equation}
where the operator $\text{hom}$ is defined in \eqref{eq:hom}.
Denote the degree of $T$ defined above in $(x_{1},x_{2})$ as $\mbox{deg}\left(T,x\right)$,
and the degree of $T$ in $\lambda$ as $\text{deg}(T,\lambda)$. 

By substituting $x_{1}$ and $x_{2}$ with $x_{1}=\frac{1-x^{2}}{1+x^{2}}$
and $x_{2}=\frac{2x}{1+x^{2}}$ respectively, let us further define
\begin{equation}
\hat{T}(l,\gamma^{2},x,\lambda)=\bar{T}(l,\gamma^{2},\frac{1-x^{2}}{1+x^{2}},\frac{2x}{1+x^{2}},\lambda),\label{eq:T_hat}
\end{equation}
and observe that the variables $x_{1},x_{2}\in\mathbb{R}$ subject
to $x_{1}^{2}+x_{2}^{2}=1$ is now parameterized by $x\in\mathbb{R}$.
It is straightforward to verify that $\left(1+x^{2}\right)^{\mbox{deg}\left(T,x\right)}\hat{T}(l,\gamma^{2},x,\lambda^{2})$
is a matrix polynomial in $(x,\lambda)$, and is a matrix homogeneous polynomial in $\lambda$.
\begin{lem}\label{le:3}
The following condition holds 
\begin{equation}
\begin{array}{c}
\begin{array}{rl}
\gamma^{2}-\overline{Q(z)\left[1-zL(z)P(z,\lambda)\right]}\cdot\\
Q(z)\left[1-zL(z)P(z,\lambda)\right] & >0
\end{array}\\
\forall z\in\mathbb{C},\left|z\right|=1,\forall\lambda\in\Lambda
\end{array}\label{eq: theorem 3 condition 1}
\end{equation}
 if and only if there exists a scalar $\epsilon>0$ and an integer
$k\in\mathbb{N}$ satisfying that 
\begin{equation}
\begin{array}{l}
\left(1+x^{2}\right)^{\mbox{deg}\left(T,x\right)}\left(\hat{T}(l,\gamma^{2},x,\lambda^{2})-\epsilon\left\Vert \lambda\right\Vert _{2}^{2\text{deg}(T,\lambda)}I\right)\left\Vert \lambda\right\Vert _{2}^{2k}\text{\ is SOS}.\end{array}\label{eq:thereom 3 condition 2}
\end{equation}

\begin{IEEEproof}
``$\Leftarrow$'' Assume that there exists a scalar $\epsilon>0$
an integer $k\in\mathbb{N}$ such that the condition \eqref{eq:thereom 3 condition 2}
is satisfied. Then, we can derive that 
\[
\begin{array}{l}
\hat{T}(l,\gamma^{2},x,\lambda^{2})>0,\forall x\in\mathbb{R},\forall\lambda\in\mathbb{R}_{0}^{n}\end{array}.
\]
Next, let us recall that $\hat{T}$ is a matrix homogeneous  polynomial
in $\lambda$ according to \eqref{eq:hom2}. Hence. it can be inferred
from Theorem 11 in \cite{chesi2010lmi} that the above condition holds if
and only if 
\[
\hat{T}(l,\gamma^{2},x,\lambda)>0,\forall x\in\mathbb{R},\forall\lambda\in\Lambda.
\]
 It follows from \eqref{eq:T_hat} that the above condition is equivalent
to 
\[
\bar{T}(l,\gamma^{2},x_{1},x_{2},\lambda)>0,\forall x_{1}^{2}+x_{2}^{2}=1,\forall\lambda\in\Lambda,
\]
 which yields that $T(l,\gamma^{2},x_{1},x_{2},\lambda)>0,\forall x_{1}^{2}+x_{2}^{2}=1,\forall\lambda\in\Lambda$.
Based on \eqref{eq:transform-1} and \eqref{eq:transform-2}, it can
verified that the condition  \eqref{eq: theorem 3 condition 1} is satisfied.

``$\Rightarrow$'' Suppose the condition in \eqref{eq: theorem 3 condition 1}
holds. It follows from the variable substitution $z=x_{1}+jx_{2}$
that \eqref{eq:transform-1} holds. Therefore, it holds that 
\[
T(l,\gamma^{2},x_{1},x_{2},\lambda)>0,\forall x_{1}^{2}+x_{2}^{2}=1,\forall\lambda\in\Lambda.
\]

Due to the compactness of $\Lambda$ and the set of unit circle of
$(x_{1},x_{2})$, there exists a small enough scalar $\epsilon>0$
such that 
\[
T(l,\gamma^{2},x_{1},x_{2},\lambda)-\epsilon I>0,\forall x_{1}^{2}+x_{2}^{2}=1,\forall\lambda\in\Lambda.
\]
Since $\sum_{i=1}^{n}\lambda_{i}=1$, one has that 
\[
\begin{array}{c}
\bar{T}(l,\gamma^{2},x_{1},x_{2},\lambda)-\epsilon\left(\sum_{i=1}^{n}\lambda_{i}\right)^{\text{deg}(T,\lambda)}I>0,\\
\forall x_{1}^{2}+x_{2}^{2}=1,\forall\lambda\in\Lambda.
\end{array}
\]

Next, let us apply the result on positivity of matrix  homogeneous 
polynomials over the simplex in Theorem 3 in \cite{scherer2006matrix}: for each fixed
pair of $(x_{1},x_{2})$, $\bar{T}(l,\gamma^{2},x_{1},x_{2},\lambda)-\epsilon\left(\sum_{i=1}^{n}\lambda_{i}\right)^{\text{deg}(T,\lambda)}I$
is a matrix homogeneous  polynomial in $\lambda$ with $\lambda$ constrained
in the simplex $\Lambda$, which follows that there exists $k(x_{1},x_{2})\in\mathbb{N}$
such that all the matrix coefficients of $\left(\bar{T}(l,\gamma^{2},x_{1},x_{2},\lambda)-\epsilon\left(\sum_{i=1}^{n}\lambda_{i}\right)^{\text{deg}(T,\lambda)}I\right)\left(\sum_{i=1}^{n}\lambda_{i}\right)^{k(x_{1},x_{2})}$
with respect to $\lambda$ are positive definite. As a consequence,
the matrix coefficients of 
\[
\begin{array}{c}
\left(\bar{T}(l,\gamma^{2},x_{1},x_{2},\lambda)-\epsilon\left(\sum_{i=1}^{n}\lambda_{i}\right)^{\text{deg}(T,\lambda)}I\right)\left(\sum_{i=1}^{n}\lambda_{i}\right)^{\hat{k}}
\end{array}
\]
with any integer $\hat{k}\ge k(x_{1},x_{2})$ can be shown to be
positive definite. Hence, it can be inferred that there exists $k\in\mathbb{N}$
such that the matrix coefficients of 
\[
\begin{array}{c}
\left(\bar{T}(l,\gamma^{2},x_{1},x_{2},\lambda)-\epsilon\left(\sum_{i=1}^{n}\lambda_{i}\right)^{\text{deg}(T,\lambda)}I\right)\left(\sum_{i=1}^{n}\lambda_{i}\right)^{k}
\end{array}
\]
with respect to $\lambda$ are positive definite for all $x_{1},x_{2}\in\mathbb{R},x_{1}^{2}+x_{2}^{2}=1.$
Let us denote $G_{d}(x_{1},x_{2})$ with $d\in\mathcal{D}\triangleq\left\{ d\in\mathbb{N}^{n}:\sum_{i=1}^{n}d_{i}=\text{deg}(T,\lambda)+k\right\} $
by the matrix coefficients, i.e., 
\begin{equation}
\begin{array}{rl}
 & \left(\bar{T}(l,\gamma^{2},x_{1},x_{2},\lambda)-\epsilon\left(\sum_{i=1}^{n}\lambda_{i}\right)^{\text{deg}(T,\lambda)}I\right)\left(\sum_{i=1}^{n}\lambda_{i}\right)^{k}\\
= & \sum_{d\in\mathcal{D}}G_{d}(x_{1},x_{2})\lambda_{1}^{d_{1}}\cdots\lambda_{n}^{d_{n}}
\end{array}\label{eq:proof_1}
\end{equation}
 wherein $G_{d}(x_{1},x_{2})>0$ for all $x_{1},x_{2}\in\mathbb{R},x_{1}^{2}+x_{2}^{2}=1.$ 

By the variable substitution $x_{1}=\frac{1-x^{2}}{1+x^{2}}$ and
$x_{2}=\frac{2x}{1+x^{2}}$, we can rewrite \eqref{eq:proof_1} equivalently
into 
\[
\begin{array}{rl}
 & \left(\hat{T}(l,\gamma^{2},x,\lambda)-\epsilon\left(\sum_{i=1}^{n}\lambda_{i}\right)^{\text{deg}(T,\lambda)}I\right)\left(\sum_{i=1}^{n}\lambda_{i}\right)^{k}\\
= & \sum_{d\in\mathcal{D}}G_{d}(\frac{1-x^{2}}{1+x^{2}},\frac{2x}{1+x^{2}})\lambda_{1}^{d_{1}}\cdots\lambda_{n}^{d_{n}}.
\end{array}
\]
Thus, $G_{d}(\frac{1-x^{2}}{1+x^{2}},\frac{2x}{1+x^{2}})>0$ for all
$x\in\mathbb{R}.$ Multiplying both sides of the above equation with
$(1+x^{2})^{\text{deg}(T,x)}$, we obtain that 
\[
\begin{array}{rl}
 & (1+x^{2})^{\text{deg}(T,x)}\left(\hat{T}(l,\gamma^{2},x,\lambda)-\epsilon\left(\sum_{i=1}^{n}\lambda_{i}\right)^{\text{deg}(T,\lambda)}I\right)\left(\sum_{i=1}^{n}\lambda_{i}\right)^{k}\\
= & \sum_{d\in\mathcal{D}}(1+x^{2})^{\text{deg}(T,x)}G_{d}(\frac{1-x^{2}}{1+x^{2}},\frac{2x}{1+x^{2}})\lambda_{1}^{d_{1}}\cdots\lambda_{n}^{d_{n}},
\end{array}
\]
and it can be inferred that $(1+x^{2})^{\text{deg}(T,x)}G_{d}(\frac{1-x^{2}}{1+x^{2}},\frac{2x}{1+x^{2}})$
is a matrix polynomial in $x$. 

Since $(1+x^{2})^{\text{deg}(T,x)}G_{d}(\frac{1-x^{2}}{1+x^{2}},\frac{2x}{1+x^{2}})$
is positive definite for all $x\in\mathbb{R},$ it can be obtained
from Theorem 4 in \cite{chesi2010lmi} that it is an SOS matrix polynomial
in $x$, i.e., it can be rewritten as
\[
(1+x^{2})^{\text{deg}(T,x)}G_{d}(\frac{1-x^{2}}{1+x^{2}},\frac{2x}{1+x^{2}})=\sum_{i}G_{di}(x)^{T}G_{di}(x)
\]
for some matrix polynomials $G_{di}$ in $x$. 

It can now be derived that 
\[
\begin{array}{rl}
 & (1+x^{2})^{\text{deg}(T,x)}\left(\hat{T}(l,\gamma^{2},x,\lambda^{2})-\epsilon\left(\sum_{i=1}^{n}\lambda_{i}^{2}\right)^{\text{deg}(T,\lambda)}I\right)\left(\sum_{i=1}^{n}\lambda_{i}^{2}\right)^{k}\\
= & \sum_{d\in\mathcal{D}}(1+x^{2})^{\text{deg}(T,x)}G_{d}(\frac{1-x^{2}}{1+x^{2}},\frac{2x}{1+x^{2}})\lambda_{1}^{2d_{1}}\cdots\lambda_{n}^{2d_{n}}\\
= & \sum_{d\in\mathcal{D}}\sum_{i}G_{di}(x)^{T}G_{di}(x)\lambda_{1}^{2d_{1}}\cdots\lambda_{n}^{2d_{n}}\\
= & \sum_{d\in\mathcal{D}}\sum_{i}\left(\lambda_{1}^{d_{1}}\cdots\lambda_{n}^{d_{n}}G_{d_i}(x)\right)^{T}\left(\lambda_{1}^{d_{1}}\cdots\lambda_{n}^{d_{n}}G_{d_i}(x)\right),
\end{array}
\]
and hence we can conclude that $(1+x^{2})^{\text{deg}(T,x)}\left(\hat{T}(l,\gamma^{2},x,\lambda^{2})-\epsilon\left\Vert \lambda\right\Vert _{2}^{2\text{deg}(T,\lambda)}I\right)\left\Vert \lambda\right\Vert _{2}^{2k}$
is SOS. 
\end{IEEEproof}
\end{lem}
\begin{rem}
We note that the result in Lemma \ref{le:3} can not be extended
trivially from Lemma \ref{lem:1} due to the existence of an extra
independent variable $z$ in \eqref{eq: theorem 3 condition 1} apart
from the uncertainty $\lambda.$ In fact, Lemma \ref{rem:3} proposes
a necessary and sufficient condition that can be verified by solving convex
optimization in the form of SDP programs for characterizing positivity of a matrix rational function
in both the standard simplex and the unit circle simultaneously. 
\end{rem}
At this point, we are ready to present the following theorem. 
\begin{thm}
\label{thm:2}The optimal solution 
$(\gamma^{*},l^{*})$ to \eqref{eq:optimzation_Z_uncertain_1} can be obtained by solving the following SOS
program

\begin{equation}
\begin{array}{rl}
\underset{\eta,\epsilon>0,k\in\mathbb{N},l}{\text{min}} & \eta\\
\text{s.t. } & \left(1+x^{2}\right)^{\mbox{deg}\left(T,x\right)}\left(\hat{T}(l,\gamma^{2},x,\lambda^{2})-\epsilon\left\Vert \lambda\right\Vert _{2}^{2\text{deg}(T,\lambda)}I\right)\left\Vert \lambda\right\Vert _{2}^{2k}\text{\ is SOS}
\end{array}\label{eq: optimization z-domain SOS-2}
\end{equation}
where $\gamma^{*}=\sqrt{\eta^{*}}$ and $l^{*}$ is given by the optimal
solution of \eqref{eq: optimization z-domain SOS-2}.\end{thm}
\begin{IEEEproof}
It can be proved directly from Lemma \ref{le:3}. 
\end{IEEEproof}
\vspace{2mm}
Theorem \ref{thm:2} provides an approach to solve Problem \ref{prob:2}
by solving a set of optimization problems following the same procedures
described in Remark \ref{rem:3}. 

\vspace{2mm}
When the case with no plant uncertainty is considered, both time-domain and frequency-domain approaches are well studied in the literature. Their results can be tied together in a clear way if both Q-filter and learning function are causal  \cite{norrlof2002time, amann1996h}.  This applies to our case with uncertain plant under consideration.  Specifically,  when $Q(q)$ and $L(q)$ are both designed
to be causal, the optimal $\gamma^{*}$ and $l^{*}$ obtained from
\eqref{eq: optimization z-domain SOS-2} will satisfy the time-domain condition $\left\Vert \mathbf{e}_{\infty}-\mathbf{e}_{j+1}\right\Vert _{2}<\gamma^{*}\left\Vert \mathbf{e}_{\infty}-\mathbf{e}_{j}\right\Vert _{2}$
for $j\in\{1,2\ldots\}$ for the ILC system 
with any finite trial length $N$. In this case, the realization of the ILC algorithm \eqref {eq: learning algorithm} in time domain can be found in  \cite{norrlof2002time,bristow2006survey}. 

\begin{rem} \label{re: BMI}
In both Problems \ref{prob:1} and \ref{prob:2}, we optimize $L(q)$ to obtain the best robust convergence rate while assuming $Q(q)$ fixed. Theorems \ref{thm:1} and \ref{thm:2} show that these can be  addressed by solving convex optimization problems. When both the Q-filter and learning function $L(q)$ are set to be decision variables (to be optimized), a suboptimal solution can be reached.
Specifically, the methods proposed in this work can be applied to solve Problem
\ref{prob:1} or \ref{prob:2} with flexible Q-filter $Q(q)$ by slight
modifications. This can be done as follows. First, choose a proper
initial $Q(q)$ and solve the SOS program in \eqref{eq:thm 1} or \eqref{eq: optimization z-domain SOS-2}.
Next, fix $L(q)$ as the obtained optimal solution, and solve the
SOS program again with variable $Q(q)$. If there is any constraint
on $Q(q)$, it can be imposed as an additional constraint in the SOS
program. Then, repeat the previous steps by iterating between $L(q)$
and $Q(q)$ to obtain a decreasing $\gamma$ as $T$ in \eqref{eq:thm 1}
or $\hat{T}$ in \eqref{eq: optimization z-domain SOS-2} is bilinear
in $L$ and $Q$. We should note that while a flexible Q-filter might benefit the transient learning behavior, the non-zero steady state tracking error $\mathbf{e}_{\infty}$ is inevitable if $Q(q)\neq1$ (see Theorem 3 in \cite{bristow2006survey}). 
\end{rem}

\section{Numerical example}
In this section, we present a numerical example to
illustrate the proposed results. The computations are done
by Matlab with the toolbox SeDuMi \cite{sturm1999using} and SOSTOOLS \cite{prajna2002introducing}.
Consider a linear uncertain plant  given by

\begin{figure} 
 \centering
\includegraphics[scale=0.5]{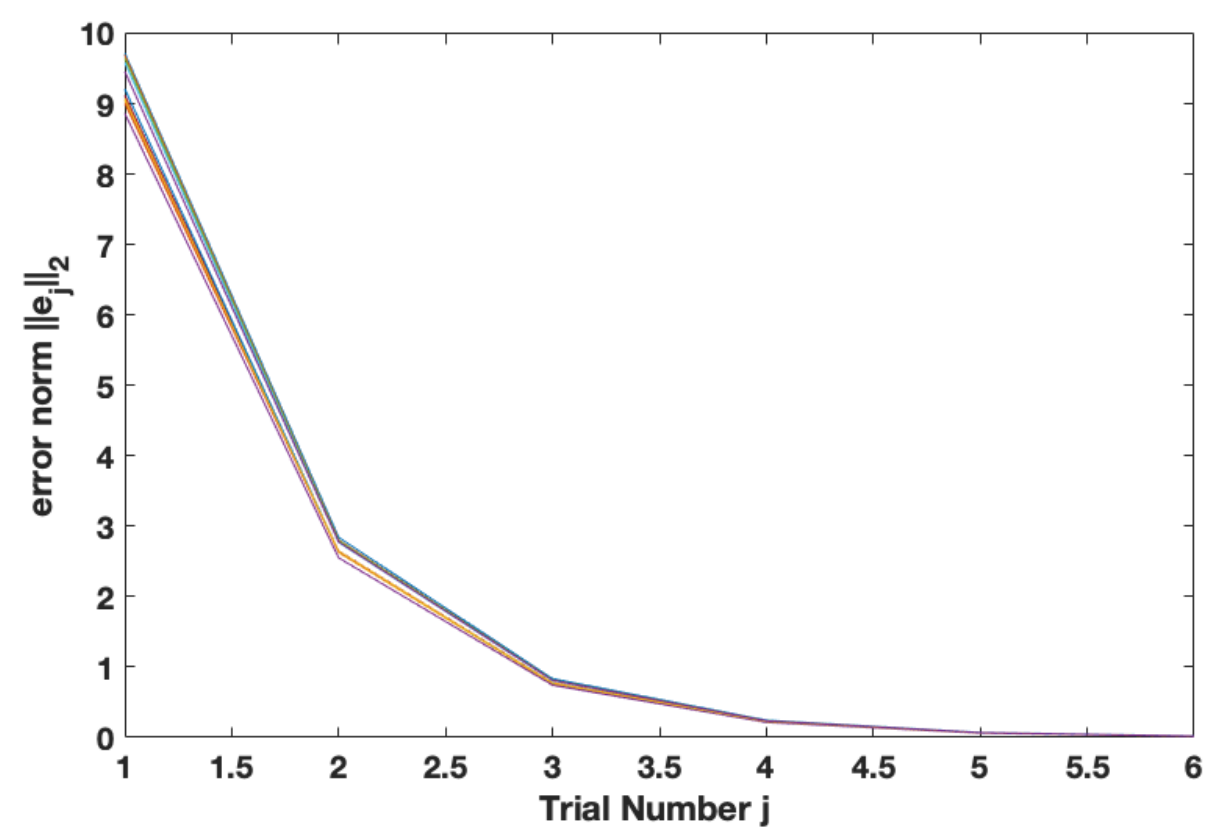}\label{fig:1}
 \centering
 \caption{Trajectories of $l_2$ norm of error signals under different values of $\theta$ and $\textbf{d}$ }
\end{figure}
\[
A=\begin{pmatrix}\theta & -0.5\\
-2\theta-0.1 & 0.2
\end{pmatrix},B=\begin{pmatrix}1\\
1
\end{pmatrix},C=\begin{pmatrix}1 & 1\end{pmatrix}
\]
where $\theta$ is an uncertainty variable that can take any value
in $\left[-0.7,-0.5\right]$. Suppose the desired output trajectory
is described by $y_{d}=\sin\left(k2\pi/N\right)$ with $N=100$, and
fix $Q(q)=1$ for simplicity. We aim at finding the optimal learning function with the
structure \eqref{eq: L(q)} or \eqref{eq:L(z)}  that minimizes the
robust convergent rate $\gamma$. Given $N=100$, one can predict high computational
complexity if the  time-domain approach in  Theorem \ref{thm:1} is adopted as $M$ in \eqref{eq:thm 1}  is a $200\times200$ matrix polynomial with high degree. 
Hence, we resort to the frequency-domain method to solve Problem \ref{prob:2}.

First of all, by exploiting the Jury stability criterion, it can be
verified that the plant is internally robust stable for all $\theta\in[-0.7,-0.5].$

Next, let us express the uncertain plant in $z$-domain as 
\[
P(z,\theta)=\frac{-40z+60\theta+16}{20z^{2}+\left(4+20\theta\right)z+16\theta+1},\theta\in[-0.7,-0.5].
\]
By utilizing the variable transform described in Remark \ref{rem:1},
the plant can be rewritten into 
\[
P(z,\lambda)=\frac{40z+30\lambda_{1}+42\lambda_{2}-16}{20z^{2}+(10\lambda_{1}+14\lambda_{2}-4)z+8\lambda_{1}+11.2\lambda_{2}-1},\lambda\in\Lambda.
\]

Then, we solve the SOS program proposed in Theorem \ref{thm:2}. 
To avoid numerical error, we set $\epsilon=0.001$. With the structure
of $L(z)=l_{0}$, $L(z)=l_{0}+l_{1}z^{-1}$ and $L(z)=l_{0}+l_{1}z^{-1}+l_{2}z^{-2}$,
we obtain the optimal solution of $\gamma^{*},l^{*}$  to \eqref{eq: optimization z-domain SOS-2} as shown in
the following table. 

\begin{center}
\begin{tabular}{|c|cc|cc|cc|}
\hline 
$k$ & $\gamma^{*}$ & $l_{0}^{*}$ & $\gamma^{*}$ & $(l_{0}^{*},l_{1}^{*})$ & $\gamma^{*}$ & $(l_{0}^{*},l_{1}^{*},l_{2}^{*})$\tabularnewline
\hline 
$0$ & 0.81 & 0.30 & 0.68 & (0.33,-0.13) & 0.46 & (0.49,0.025,0.31)\tabularnewline
\hline 
$1$ & 0.81 & 0.30 & 0.68 & (0.33,-0.13) & 0.46 & (0.49,0.027,0.31)\tabularnewline
\hline 
$2$ & 0.81 & 0.30 & 0.68 & (0.33,-0.13) & 0.46 & (0.49,0.027,0.31)\tabularnewline
\hline 
$3$ & 0.81 & 0.30 & 0.68 & (0.33,-0.13) & 0.46 & (0.49, 0.027,0.31)\tabularnewline
\hline 
\end{tabular}
\end{center}
As it can be seen from the table, the additional variable in $L(z)$
gives a boost to the performance. As mentioned in Remark \ref{re: BMI}, one can improve the convergence rate $\gamma$ by utilizing a non-identity Q-filter. For instance, if we take the example with $L(z)=l_0$ and $Q(z)=1+q_{1}z^{-1}$, we first fix $Q(z)=1$ ($q_{1}=0$), and according to the above table, it is obtained that $\gamma^{*}=0.81$ associated with $l_0^{*}=0.30$. Next, one can fix $L(z)=l_0^{*}=0.30$ and solve the optimization in Theorem \ref{thm:2} again with decision variable changed from $l_0$ to $q_{1}$. Then, it can be obtained that $\gamma^{*}=0.67$ and $q_1^{*}=0.32$. We note that although the convergence rate $\gamma$ is improved by exploiting the non-identity $Q(z)=1+0.32z^{-1}$, the tracking error does not converge to zero.

When the learning function of the form $L(z)=l_{0}+l_{1}z^{-1}+l_{2}z^{-2}+l_{3}z^{-3}$
is considered, by solving the SOS program in Theorem \ref{thm:2} with $Q(z)=1$,
we obtain that $\gamma^{*}=0.32$ and the optimal leaning function
is 
\[
(l_{0}^{*},l_{1}^{*},l_{2}^{*},l_{3}^{*})=(0.51,-0.072,0.19,-0.20).
\]

Since $Q(q)$ and $L(q)$ are designed to be causal, the convergence rate obtained in z-domain holds also for that in time domain for any finite trial length $N$, i.e., $\|\mathbf{e}_{\infty} -\mathbf{e}_{j+1} \|_{2}<0.32 \|\mathbf{e}_{\infty} -\mathbf{e}_{j} \|_{2}$. 
In the end, by exploiting the obtained $L^{*}(q)$, we simulate the iterative learning process \eqref{eq:uncertain plant 3}, \eqref{eq:error} and \eqref{eq: learning in time}  in time domain.  Since $Q(z)=1$, it follows from \cite{phan2000unified} that $\mathbf{e_{\infty}}=0$, and it is expected that $\| \mathbf{e}_{j+1} \|_{2}<0.32 \|\mathbf{e}_{j} \|_{2}$.
 With randomly values of $\theta\in[-0.7,-0.5]$, and randomly generated
disturbance $\mathbf{d}$, the trajectories for the $l_{2}$ norm
of the error $\mathbf{e_{j}}$ are depicted in Figure 1. As shown in Figure 1, the ILC system is robustly monotonically convergent.

\section{Conclusion}

This work has considered robust monotonic convergent ILC for uncertain
linear time-invariant systems. Both time domain and frequency domain
are discussed. To establish whether the ILC system is monotonically
convergent, a necessary and sufficient condition in the form of SOS
program has been provided. This condition provides us with an approach
to optimize the convergence speed by solving a set of convex optimization
problems. 
Inspired by \cite{van2009iterative, ardakani2017convergence},  an interesting future direction is to further develop non-causal ILC algorithms in frequency domain while removing the assumption on  finite trial interval. 
\bibliographystyle{IEEEtran}
\bibliography{ILC_reference}

\end{document}